%% file: main.tex
\pdfoutput=1

\documentclass[12pt,a4paper]{article}

\usepackage{ifthen}
\newboolean{pdflatex}
\setboolean{pdflatex}{true}

\newboolean{articletitles}
\setboolean{articletitles}{true}

\newboolean{uprightparticles}
\setboolean{uprightparticles}{false}

\def\paperauthors{LHCb collaboration}
\def\paperasciititle{Observation of the Bs to Dstar D decay}
\def\papertitle{Observation\\ of the $B_s^0 \rightarrow D^{\ast \pm} D^{\mp}$ decay}
\def\paperkeywords{{High Energy Physics}, {LHCb}}
\def\papercopyright{\the\year\ CERN for the benefit of the LHCb collaboration}
\def\paperlicence{CC BY 4.0 licence}
\def\paperlicenceurl{https://creativecommons.org/licenses/by/4.0/}

\input{preamble}
\input{owncommands}
\usepackage{longtable}
\usepackage{multirow}
\usepackage{booktabs}

\begin{document}

\renewcommand{\thefootnote}{\fnsymbol{footnote}}
\setcounter{footnote}{1}

\input{title-LHCb-PAPER}

\renewcommand{\thefootnote}{\arabic{footnote}}
\setcounter{footnote}{0}

\pagestyle{plain}
\pagenumbering{arabic}

\input{text}

\input{acknowledgements}


\addcontentsline{toc}{section}{References}
\bibliographystyle{LHCb}
\bibliography{main,standard,LHCb-PAPER,LHCb-CONF,LHCb-DP,LHCb-TDR}

\newpage
\input{LHCb_Authorship_22-Sep-2020}

\end{document}

%% file: preamble.tex
\usepackage[top=1in, bottom=1.25in, left=1in, right=1in]{geometry}

\usepackage{microtype}
\usepackage{lineno}  
\usepackage{xspace} 
\usepackage{caption} 

\usepackage{graphicx}  
\usepackage{color}
\usepackage{colortbl}
\graphicspath{{./figs/}} 

\usepackage{amsmath} 
\usepackage{amssymb}
\usepackage{amsfonts}
\usepackage{upgreek} 

\newcommand*\patchAmsMathEnvironmentForLineno[1]{%
\expandafter\let\csname old#1\expandafter\endcsname\csname #1\endcsname
\expandafter\let\csname oldend#1\expandafter\endcsname\csname
end#1\endcsname
 \renewenvironment{#1}%
   {\linenomath\csname old#1\endcsname}%
   {\csname oldend#1\endcsname\endlinenomath}%
}
\newcommand*\patchBothAmsMathEnvironmentsForLineno[1]{%
  \patchAmsMathEnvironmentForLineno{#1}%
  \patchAmsMathEnvironmentForLineno{#1*}%
}
\AtBeginDocument{%
\patchBothAmsMathEnvironmentsForLineno{equation}%
\patchBothAmsMathEnvironmentsForLineno{align}%
\patchBothAmsMathEnvironmentsForLineno{flalign}%
\patchBothAmsMathEnvironmentsForLineno{alignat}%
\patchBothAmsMathEnvironmentsForLineno{gather}%
\patchBothAmsMathEnvironmentsForLineno{multline}%
\patchBothAmsMathEnvironmentsForLineno{eqnarray}%
}

\usepackage{hyperxmp}

\usepackage[pdftex,
            pdfauthor={\paperauthors},
            pdftitle={\paperasciititle},
            pdfkeywords={\paperkeywords},
            pdfcopyright={Copyright (C) \papercopyright},
            pdflicenseurl={\paperlicenceurl}]{hyperref}

\usepackage[colorinlistoftodos,textsize=scriptsize]{todonotes}

\usepackage[bottom,flushmargin,hang,multiple]{footmisc}

\usepackage[all]{hypcap} 

\input{lhcb-symbols-def} 

\usepackage{cite} 
\usepackage{mciteplus}

%% file: lhcb-symbols-def.tex
\usepackage{xspace} 
\usepackage{upgreek}

\def\lhcb   {\mbox{LHCb}\xspace}

\def\MagUp {\mbox{\em Mag\kern -0.05em Up}\xspace}

\ifthenelse{\boolean{uprightparticles}}%
{

 \def\Ppi         {\ensuremath{\uppi}\xspace}

 \def\Pphi        {\ensuremath{\upphi}\xspace}

 \def\PDelta      {\ensuremath{\Delta}\xspace}                 
 \def\PXi         {\ensuremath{\Xi}\xspace}                 
 \def\PLambda     {\ensuremath{\Lambda}\xspace}                 
 \def\PSigma      {\ensuremath{\Sigma}\xspace}                 
 \def\POmega      {\ensuremath{\Omega}\xspace}                 
 \def\PUpsilon    {\ensuremath{\Upsilon}\xspace}

 \def\PB      {\ensuremath{\mathrm{B}}\xspace}                 
                  
 \def\PD      {\ensuremath{\mathrm{D}}\xspace}

 \def\PK      {\ensuremath{\mathrm{K}}\xspace}

 \def\PW      {\ensuremath{\mathrm{W}}\xspace}

 \def\Pb      {\ensuremath{\mathrm{b}}\xspace}                 
 \def\Pc      {\ensuremath{\mathrm{c}}\xspace}                 
 \def\Pd      {\ensuremath{\mathrm{d}}\xspace}

 \def\Pi      {\ensuremath{\mathrm{i}}\xspace}

 \def\Pp      {\ensuremath{\mathrm{p}}\xspace}

 \def\Ps      {\ensuremath{\mathrm{s}}\xspace}

 \def\thebaroffset{0.0em}
}
{

 \def\Ppi         {\ensuremath{\pi}\xspace}

 \def\Pphi        {\ensuremath{\phi}\xspace}

 \mathchardef\PDelta="7101
 \mathchardef\PXi="7104
 \mathchardef\PLambda="7103
 \mathchardef\PSigma="7106
 \mathchardef\POmega="710A
 \mathchardef\PUpsilon="7107
                  
 \def\PB      {\ensuremath{B}\xspace}                 
                  
 \def\PD      {\ensuremath{D}\xspace}

 \def\PK      {\ensuremath{K}\xspace}

 \def\PW      {\ensuremath{W}\xspace}

 \def\Pb      {\ensuremath{b}\xspace}                 
 \def\Pc      {\ensuremath{c}\xspace}                 
 \def\Pd      {\ensuremath{d}\xspace}

 \def\Pi      {\ensuremath{i}\xspace}

 \def\Pp      {\ensuremath{p}\xspace}

 \def\Ps      {\ensuremath{s}\xspace}

 \def\thebaroffset{0.18em}
}
\newcommand{\offsetoverline}[2][\thebaroffset]{\kern #1\overline{\kern -#1 #2}}%

\makeatletter
\ifcase \@ptsize \relax
  \newcommand{\miniscule}{\@setfontsize\miniscule{4}{5}}
\or
  \newcommand{\miniscule}{\@setfontsize\miniscule{5}{6}}
\or
  \newcommand{\miniscule}{\@setfontsize\miniscule{5}{6}}
\fi
\makeatother

\DeclareRobustCommand{\optbar}[1]{\shortstack{{\miniscule (\rule[.5ex]{1.25em}{.18mm})}
  \\ [-.7ex] $#1$}}

\def\W      {{\ensuremath{\PW}}\xspace}

\def\dquark    {{\ensuremath{\Pd}}\xspace}

\def\squark    {{\ensuremath{\Ps}}\xspace}

\def\cquark    {{\ensuremath{\Pc}}\xspace}

\def\bquark    {{\ensuremath{\Pb}}\xspace}

\def\pion   {{\ensuremath{\Ppi}}\xspace}
\def\piz    {{\ensuremath{\pion^0}}\xspace}
\def\pip    {{\ensuremath{\pion^+}}\xspace}
\def\pim    {{\ensuremath{\pion^-}}\xspace}

\def\kaon    {{\ensuremath{\PK}}\xspace}

\def\KorKbar {\kern \thebaroffset\optbar{\kern -\thebaroffset \PK}{}\xspace}

\def\Kp      {{\ensuremath{\kaon^+}}\xspace}
\def\Km      {{\ensuremath{\kaon^-}}\xspace}

\def\D       {{\ensuremath{\PD}}\xspace}

\def\DorDbar {\kern \thebaroffset\optbar{\kern -\thebaroffset \PD}\xspace}
\def\Dz      {{\ensuremath{\D^0}}\xspace}

\def\Dp      {{\ensuremath{\D^+}}\xspace}
\def\Dm      {{\ensuremath{\D^-}}\xspace}

\def\Dmp     {{\ensuremath{\D^\mp}}\xspace}
\def\DpDm    {\ensuremath{\Dp {\kern -0.16em \Dm}}\xspace}
\def\Dstar   {{\ensuremath{\D^*}}\xspace}

\def\Dstarp  {{\ensuremath{\D^{*+}}}\xspace}
\def\Dstarm  {{\ensuremath{\D^{*-}}}\xspace}
\def\Dstarpm {{\ensuremath{\D^{*\pm}}}\xspace}
\def\Dstarmp {{\ensuremath{\D^{*\mp}}}\xspace}

\def\Dsp     {{\ensuremath{\D^+_\squark}}\xspace}

\def\Dsmp    {{\ensuremath{\D^{\mp}_\squark}}\xspace}

\def\B       {{\ensuremath{\PB}}\xspace}
\def\Bbar    {{\ensuremath{\offsetoverline{\PB}}}\xspace}

\def\BorBbar {\kern \thebaroffset\optbar{\kern -\thebaroffset \PB}\xspace}
\def\Bz      {{\ensuremath{\B^0}}\xspace}

\def\Bd      {{\ensuremath{\B^0}}\xspace}
\def\Bdb     {{\ensuremath{\Bbar{}^0}}\xspace}
\def\BdorBdbar {\kern \thebaroffset\optbar{\kern -\thebaroffset \Bd}\xspace}

\def\Bs      {{\ensuremath{\B^0_\squark}}\xspace}
\def\Bsb     {{\ensuremath{\Bbar{}^0_\squark}}\xspace}
\def\BsorBsbar {\kern \thebaroffset\optbar{\kern -\thebaroffset \Bs}\xspace}

\def\Bds     {{\ensuremath{\B_{(\squark)}^0}}\xspace}

\def\Y#1S{\ensuremath{\PUpsilon{(#1S)}}\xspace}

\def\proton      {{\ensuremath{\Pp}}\xspace}

\def\Lz          {{\ensuremath{\PLambda}}\xspace}

\def\LorLbar     {\kern \thebaroffset\optbar{\kern -\thebaroffset \PLambda}\xspace}

\def\Lc          {{\ensuremath{\Lz^+_\cquark}}\xspace}

\def\Lb           {{\ensuremath{\Lz^0_\bquark}}\xspace}

\def\BF         {{\ensuremath{\mathcal{B}}}\xspace}

\newcommand{\decay}[2]{\ensuremath{#1\!\to #2}\xspace} 

\def\to                 {\ensuremath{\rightarrow}\xspace}

\def\CP                {{\ensuremath{C\!P}}\xspace}

\newcommand{\DGd}{{\ensuremath{\Delta\Gamma_{\dquark}}}\xspace}

\newcommand{\phis}{{\ensuremath{\phi_{\squark}}}\xspace}

\def\AT#1     {\ensuremath{A_{\mathrm{T}}^{#1}}\xspace}           

\def\C#1      {\ensuremath{\mathcal{C}_{#1}}\xspace}                       
\def\Cp#1     {\ensuremath{\mathcal{C}_{#1}^{'}}\xspace}                    
\def\Ceff#1   {\ensuremath{\mathcal{C}_{#1}^{\mathrm{(eff)}}}\xspace}        
\def\Cpeff#1  {\ensuremath{\mathcal{C}_{#1}^{'\mathrm{(eff)}}}\xspace}       
\def\Ope#1    {\ensuremath{\mathcal{O}_{#1}}\xspace}                       
\def\Opep#1   {\ensuremath{\mathcal{O}_{#1}^{'}}\xspace}                    

\newcommand{\nospaceunit}[1]{\ensuremath{\text{#1}}}       
\newcommand{\aunit}[1]{\ensuremath{\text{\,#1}}}       

\newcommand{\tev}{\aunit{Te\kern -0.1em V}\xspace}
\newcommand{\gev}{\aunit{Ge\kern -0.1em V}\xspace}
\newcommand{\mev}{\aunit{Me\kern -0.1em V}\xspace}
\newcommand{\kev}{\aunit{ke\kern -0.1em V}\xspace}
\newcommand{\ev}{\aunit{e\kern -0.1em V}\xspace}
 
\newcommand{\mevc}{\ensuremath{\aunit{Me\kern -0.1em V\!/}c}\xspace}
\newcommand{\gevc}{\ensuremath{\aunit{Ge\kern -0.1em V\!/}c}\xspace}
\newcommand{\mevcc}{\ensuremath{\aunit{Me\kern -0.1em V\!/}c^2}\xspace}
\newcommand{\gevcc}{\ensuremath{\aunit{Ge\kern -0.1em V\!/}c^2}\xspace}

\def\mum  {\ensuremath{\,\upmu\nospaceunit{m}}\xspace}

\def\fb   {\ensuremath{\aunit{fb}}\xspace}
\def\invfb   {\ensuremath{\fb^{-1}}\xspace}

\def\ps   {\ensuremath{\aunit{ps}}\xspace}

\newcommand{\chisq}{\ensuremath{\chi^2}\xspace}

\newcommand{\chisqip}{\ensuremath{\chi^2_{\text{IP}}}\xspace}

\def\gsim{{~\raise.15em\hbox{$>$}\kern-.85em
          \lower.35em\hbox{$\sim$}~}\xspace}
\def\lsim{{~\raise.15em\hbox{$<$}\kern-.85em
          \lower.35em\hbox{$\sim$}~}\xspace}

\def\pt         {\ensuremath{p_{\mathrm{T}}}\xspace}

\def\ptot       {\ensuremath{p}\xspace}

\def\evtgen     {\mbox{\textsc{EvtGen}}\xspace}

\def\geant      {\mbox{\textsc{Geant4}}\xspace}

\def\photos     {\mbox{\textsc{Photos}}\xspace}

\def\pythia     {\mbox{\textsc{Pythia}}\xspace}

\def\tell1  {TELL1\xspace}
\def\ukl1   {UKL1\xspace}

\newcommand{\eg}{\mbox{\itshape e.g.}\xspace}
\newcommand{\ie}{\mbox{\itshape i.e.}\xspace}



%% file: owncommands.tex
\newcommand{\decayt}[2]{\ensuremath{#1\!\to #2}\xspace}
\renewcommand{\decay}[2]{\mbox{\ensuremath{#1\!\to #2}}\xspace}

\newcommand {\BstoDstD} {\ensuremath{\decayt{\Bs}{\Dstarpm\Dmp}}\xspace}
\newcommand {\BtoDstD} {\ensuremath{\decayt{\Bd}{\Dstarpm\Dmp}}\xspace}

\def\Dsstarpm    {{\ensuremath{\D^{*\pm}_\squark}}\xspace}
\def\Bsd         {\ensuremath{\B^{0}_{(\squark)}}\xspace}

%% file: title-LHCb-PAPER.tex
\begin{titlepage}
\pagenumbering{roman}

\vspace*{-1.5cm}
\centerline{\large EUROPEAN ORGANIZATION FOR NUCLEAR RESEARCH (CERN)}
\vspace*{1.5cm}
\noindent
\begin{tabular*}{\linewidth}{lc@{\extracolsep{\fill}}r@{\extracolsep{0pt}}}
\ifthenelse{\boolean{pdflatex}}
{\vspace*{-1.5cm}\mbox{\!\!\!\includegraphics[width=.14\textwidth]{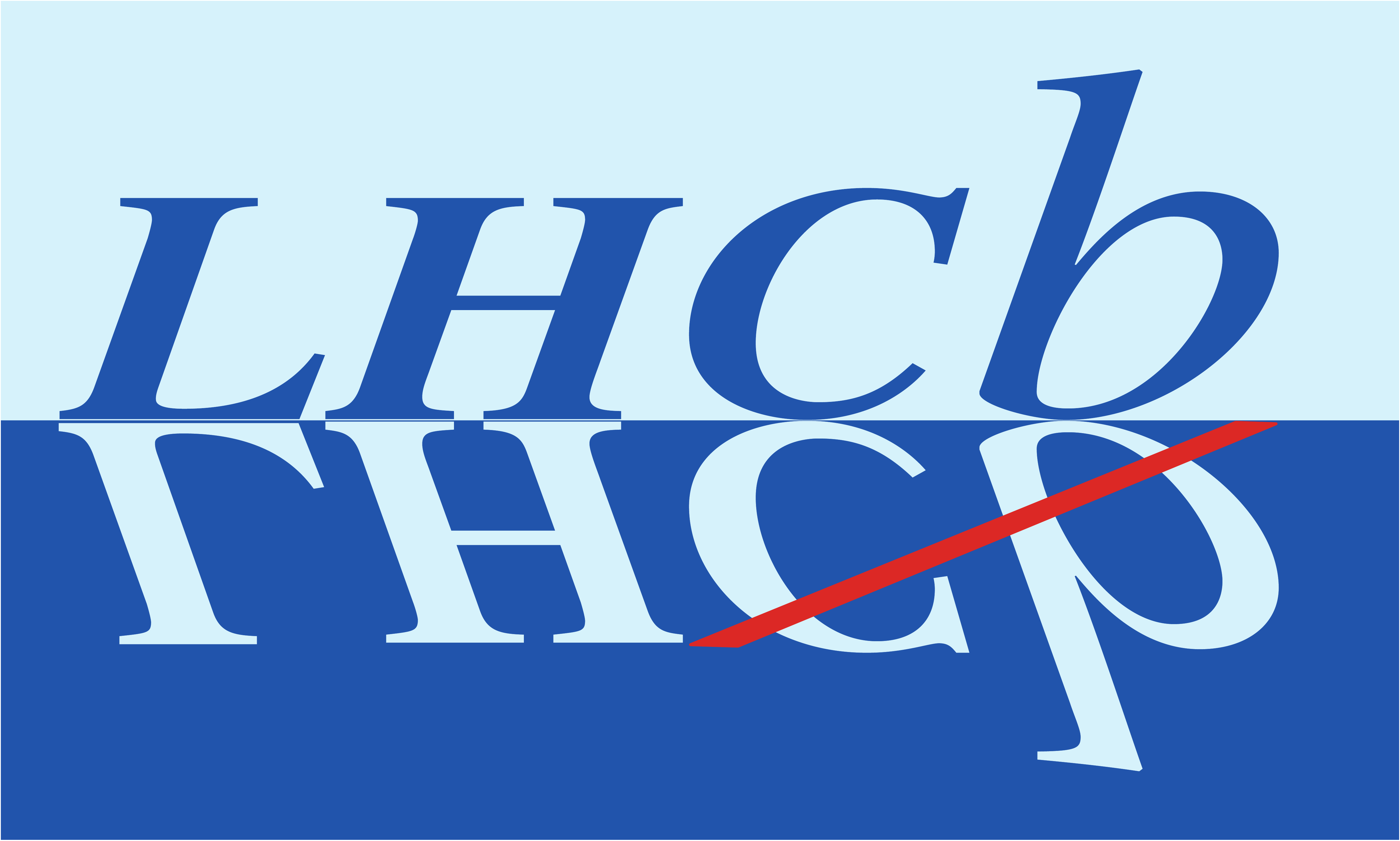}} & &}
{\vspace*{-1.2cm}\mbox{\!\!\!\includegraphics[width=.12\textwidth]{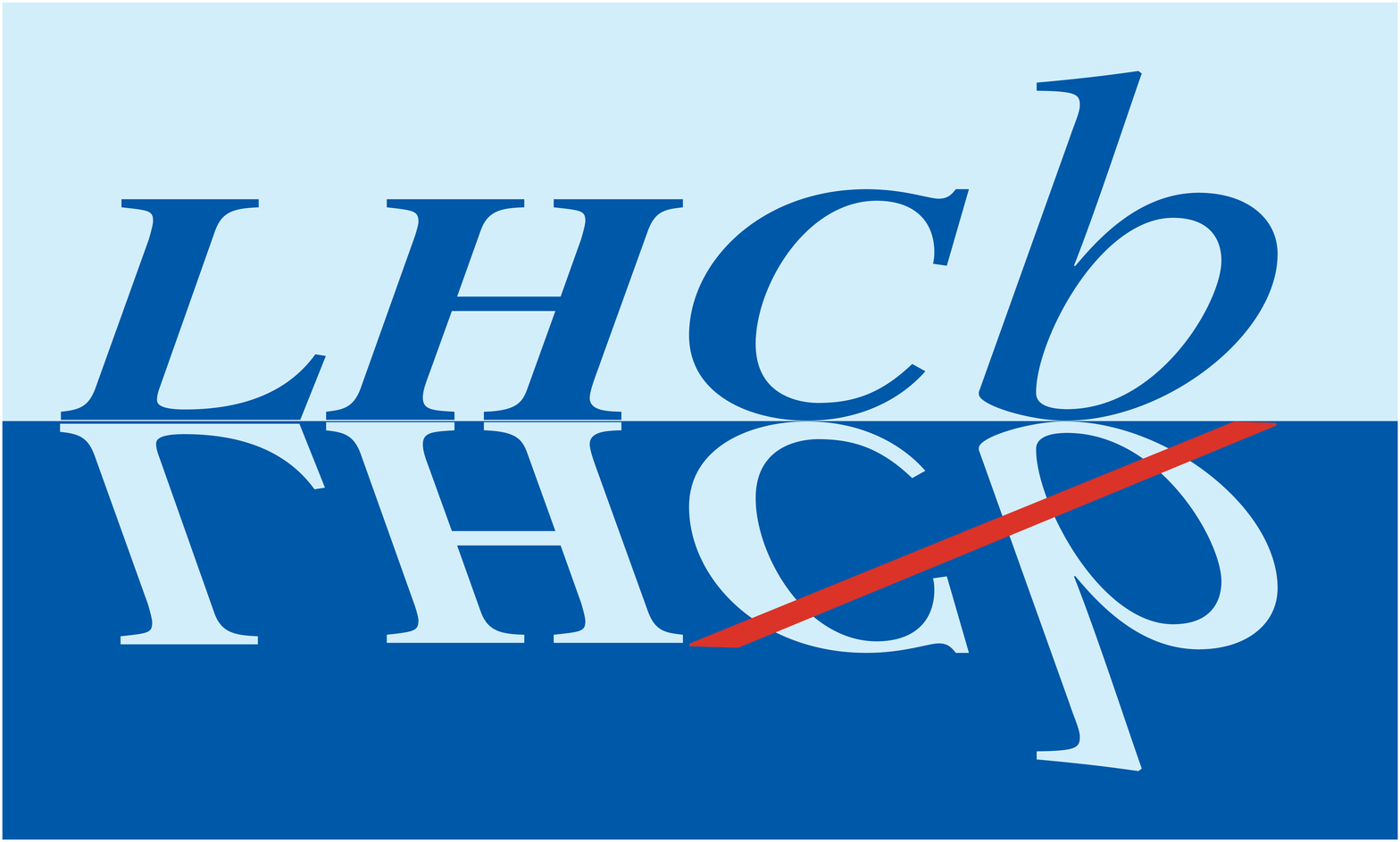}} & &}
\\
 & & CERN-EP-2020-230 \\
 & & LHCb-PAPER-2020-037 \\
 & & March 18, 2021 \\ 
 & & \\
\end{tabular*}

\vspace*{4.0cm}

{\normalfont\bfseries\boldmath\huge
\begin{center}
  \papertitle 
\end{center}
}

\vspace*{2.0cm}

\begin{center}
\paperauthors\footnote{Authors are listed at the end of this paper.}
\end{center}

\vspace{\fill}

\begin{abstract}
  \noindent
  A search for the \ensuremath{{\ensuremath{B^0_\squark}}\xspace\!\rightarrow {\ensuremath{D^{*\pm}}}\xspace {\ensuremath{D^\mp}}\xspace}\xspace decay is performed using proton-proton collision data at centre-of-mass energies of 7, 8 and 13\,TeV collected by the LHCb experiment, corresponding to an integrated luminosity of $9\ensuremath{\fb^{-1}}\xspace$.
  The decay is observed with a high significance and its
  branching fraction relative to the \ensuremath{{\ensuremath{B^0}}\xspace\!\rightarrow {\ensuremath{D^{*\pm}}}\xspace {\ensuremath{D^\mp}}\xspace}\xspace decay is measured to be
  \begin{align*}
    \frac{\mathcal{B}(B_s^0 \rightarrow D^{\ast \pm} D^{\mp}) }{\mathcal{B}(B^0 \rightarrow D^{\ast \pm} D^{\mp}) } = 0.137 \pm 0.017 \pm 0.002 \pm 0.006 \,,
  \end{align*}
  where the first uncertainty is statistical, the second systematic and the third is due to the uncertainty on the ratio of the $B_s^0$ and $B^0$  hadronisation fractions.
\end{abstract}

\vspace*{2.0cm}

\begin{center}
  Published in JHEP 03 (2021) 099
\end{center}

\vspace{\fill}

{\footnotesize
\centerline{\copyright~\papercopyright. \href{\paperlicenceurl}{\paperlicence}.}}
\vspace*{2mm}

\end{titlepage}

\newpage
\setcounter{page}{2}
\mbox{~}

%% file: text.tex
\section{Introduction}
\label{sec:Introduction}

The family of \B-meson decays into a pair of open-charm mesons
are sensitive to elements of the Cabibbo–Kobayashi–Maskawa matrix~\cite{Cabibbo:1963yz,Kobayashi:1973fv}.
While \decay{\Bd}{\D^{(*)+}\D^{(*)-}} decays can be used to measure the \Bd-\Bdb mixing phase, $\sin(2\beta)$~\cite{Aubert:2006ia,Aubert:2008ah,Fratina:2007zk,Rohrken:2012ta,LHCb-PAPER-2016-037,LHCb-PAPER-2019-036}, \decay{\Bs}{\D_s^{(*)+}\D_s^{(*)-}} decays provide access to the \Bs-\Bsb mixing phase, \phis~\cite{LHCb-PAPER-2014-051}.
Information on additional decays, such as \BstoDstD, can be exploited to constrain loop and non-factorisable contributions~\cite{ALEKSAN1993173,Sanda:1996pm,Xing:1998ca,Xing:1999yx,Pham:1999fy,Jung:2014jfa}, which can be notably prominent~\cite{Bel:2015wha}.

\begin{figure}[b]
\centering
\includegraphics[width=0.49\textwidth]{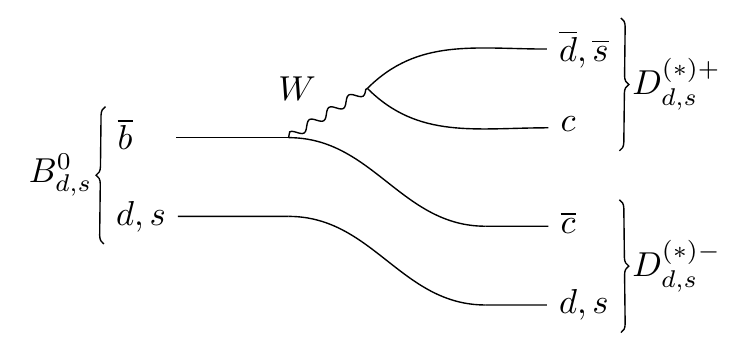}
\includegraphics[width=0.49\textwidth]{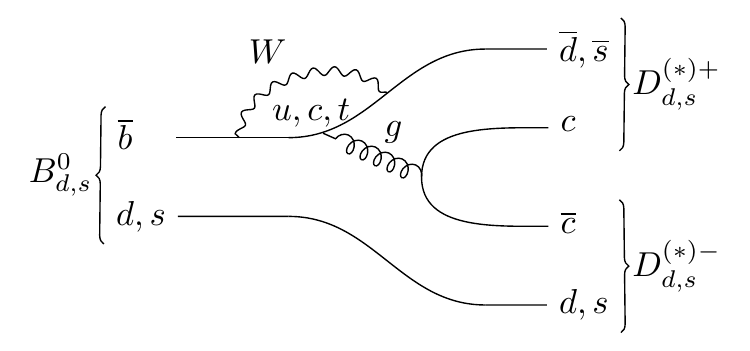}
\caption{(Left) Tree-level and (right) penguin diagrams contributing to \decay{\Bd}{\D_{(s)}^{(*)\pm}\D^{(*)\mp}} and \decay{\Bs}{\D_{(s)}^{(*)\pm} \D_{s}^{(*)\mp}} decays.}
\label{fig:feynmandiagrams1}
\end{figure}

\begin{figure}[tb]
\centering
\includegraphics[width=0.49\textwidth]{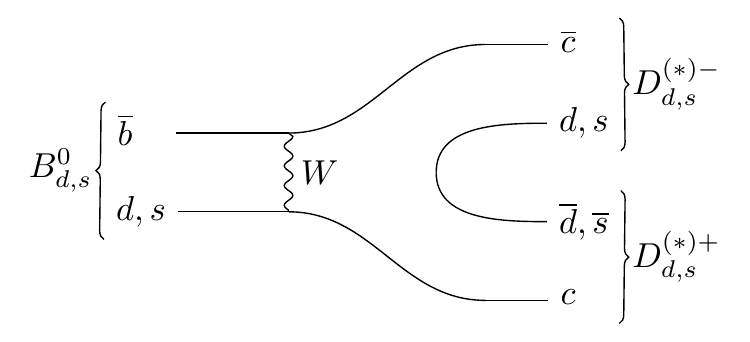}
\includegraphics[width=0.49\textwidth]{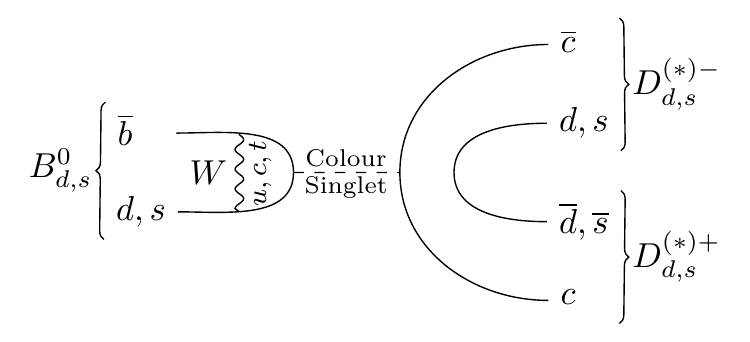}
\caption{(Left) \W-exchange and (right) penguin-annihilation diagrams contributing to \decay{\B_{(s)}^{0}}{\D_{(s)}^{(*)+}\D_{(s)}^{(*)-}} decays.}
\label{fig:feynmandiagrams2}
\end{figure}

Both \decay{\Bd}{\D^{(*)+}\D^{(*)-}} and \decay{\Bs}{\D_s^{(*)+}\D_s^{(*)-}} decays occur predominantly through tree or penguin transitions, as shown in Fig.~\ref{fig:feynmandiagrams1}.
Subleading contributions are expected from \W-exchange and penguin-annihilation transitions, illustrated in Fig.~\ref{fig:feynmandiagrams2}.
In contrast, the \BstoDstD decay is forbidden at tree level and its dominant contributions originate from \W-exchange and penguin-annihilation diagrams shown in Fig.~\ref{fig:feynmandiagrams2}, or from rescattering of intermediate states~\cite{Gronau:2012gs}.
Thus, the \BstoDstD decay can be used to estimate the subleading contributions of the \BtoDstD decay mode.

The \BstoDstD decay has not been previously observed, but an excess of possible \BstoDstD candidates was seen in a recent measurement of \CP violation in \mbox{\BtoDstD} decays by the LHCb experiment~\cite{LHCb-PAPER-2019-036}.
Assuming prominent contributions from rescattering of \eg\ \Dsstarpm\Dsmp states, the branching fraction is predicted to be $(6.1\pm3.6)\times 10^{-5}$~\cite{Gronau:2012gs}.
A perturbative QCD approach predicts a much larger branching fraction of $(3.6\pm0.6)\times 10^{-3}$~\cite{Li:2003az}.

This paper presents the first observation of the \decay{\Bs}{\Dstarp\Dm} and \decay{\Bs}{\Dstarm\Dp} decays, which have indistinguishable final states.
Throughout this paper these decays are treated together and charge conjugation is applied.
The branching fraction of the \BstoDstD decay is measured relative to the \BtoDstD decay.
Since both decay channels have the same final state, the experimental systematic uncertainties on the ratio of branching fractions are expected to be small.
The measurement uses proton-proton ($pp$) collision data collected with the LHCb detector in the years 2011, 2012, and 2015--2018 at centre-of-mass energies of 7, 8, and 13\tev, respectively, corresponding to an integrated luminosity of $9\invfb$.

\section{Detector and simulation}
\label{sec:Detector}
The \lhcb detector~\cite{LHCb-DP-2008-001,LHCb-DP-2014-002} is a single-arm forward
spectrometer covering the \mbox{pseudorapidity} range $2<\eta <5$,
designed for the study of particles containing \bquark or \cquark
quarks. The detector includes a high-precision tracking system
consisting of a silicon-strip vertex detector surrounding the $pp$
interaction region, a large-area silicon-strip detector located
upstream of a dipole magnet with a bending power of about
$4{\mathrm{\,Tm}}$, and three stations of silicon-strip detectors and straw
drift tubes placed downstream of the magnet.
The tracking system provides a measurement of the momentum, \ptot, of charged particles with
a relative uncertainty that varies from 0.5\% at low momentum to 1.0\% at 200\gevc.
The minimum distance of a track to a primary $pp$ collision vertex (PV), the impact parameter (IP),
is measured with a resolution of $(15+29/\pt)\mum$,
where \pt is the component of the momentum transverse to the beam, in\,\gevc.
Different types of charged hadrons are distinguished using information
from two ring-imaging Cherenkov detectors.
Photons, electrons and hadrons are identified by a calorimeter system consisting of
scintillating-pad and preshower detectors, an electromagnetic
and a hadronic calorimeter. Muons are identified by a
system composed of alternating layers of iron and multiwire
proportional chambers.

Simulated data samples are used to train a multivariate algorithm, model the shapes of mass distributions and calculate efficiencies.
In the simulation, $pp$ collisions are generated using
\pythia~\cite{Sjostrand:2007gs,*Sjostrand:2006za} with a specific \lhcb configuration~\cite{LHCb-PROC-2010-056}.
Decays of unstable particles
are described by \evtgen~\cite{Lange:2001uf}, in which final-state
radiation is generated using \photos~\cite{Golonka:2005pn}.
The interaction of the generated particles with the detector, and its response,
are implemented using the \geant
toolkit~\cite{Allison:2006ve, *Agostinelli:2002hh} as described in
Ref.~\cite{LHCb-PROC-2011-006}.

The distributions of particle identification (PID) variables do not match perfectly between simulation and data.
This difference is corrected using an approach where functions are constructed that transform the simulated PID response to match calibration samples of recorded data.
This is based on a four-dimensional kernel density estimation for distributions in PID value, $p_\text{T}$ and $\eta$ of the track and the event multiplicity~\cite{Poluektov:2014rxa}.

\section{Candidate selection}\label{sec:Selection}

Due to varying data-taking conditions, the data samples for the three periods 2011--2012, 2015--2016 and 2017--2018 are treated differently.
The online event selection is performed by a trigger,
which consists of a hardware stage, based on information from the calorimeter and muon
systems, followed by a software stage, which applies a full event
reconstruction.
At the hardware trigger stage, events are required to have a muon with high \pt or a
hadron, photon or electron with high transverse energy in the calorimeters.
The software trigger requires a two-, three- or four-track
secondary vertex with a significant displacement from any PV.
At least one charged particle must have a large transverse momentum and be inconsistent with originating from any PV.
A multivariate algorithm~\cite{BBDT,LHCb-PROC-2015-018} is used for
the identification of secondary vertices consistent with the decay
of a \bquark hadron.

The \decay{\Bsd}{\Dstarpm\Dmp} candidates are reconstructed through the decays \decay{\Dstarp}{\Dz\pip} with \decay{\Dz}{\Km\pip} and \decay{\Dm}{\Kp\pim\pim}.
The tracks of the final-state particles are required to have a good quality, fulfil loose PID criteria, and have a high \chisqip\ value with respect to any PV, where \chisqip is defined as the difference in the vertex-fit \chisq of a given PV reconstructed with and without the particle being considered.
The probability of a candidate being a duplicate track is required to be small.
Additionally, the distance of closest approach between all possible combinations of tracks is required to be small.
The reconstructed masses of the \Dstarp, \Dz and \Dm candidates are required to lie inside a mass window of $\pm50\mevcc$ around their known values \cite{PDG2020}, and the difference of the reconstructed masses between the \Dstarp and \Dz candidates is required to be smaller than $150\mevcc$.
The ratio of the \Dm decay time and its uncertainty, $t/\sigma_t$, is required to be larger than $-1$.
The \Bsd candidate is reconstructed by combining the \Dstarpm and \Dmp candidates to form a common vertex.
In case multiple PVs are reconstructed in the same event, the PV for which the \Bsd candidate has the lowest \chisqip is assigned as the associated PV.
The sum of the transverse momenta of the decay products of the \Bsd candidate is required to be larger than $5\gevc$ and the \chisqip of the \Bsd candidate for the associated PV is required to be small.
Furthermore, the decay time of the \Bsd candidate is required to be larger than $0.2\ps$.
Candidates are retained if the mass of the \Dstarpm\Dmp system, $m_{\Dstarpm\Dmp}$, is in the range $5000\mevcc<m_{\Dstarpm\Dmp}<5600\mevcc$.

Background contributions to \Dp candidates arise when kaons or protons stemming from hadronic decays of \Dsp and \Lc hadrons are misidentified as pions.
A combination of mass and PID requirements is used to suppress contributions from \decay{\Bd}{\Dstarm \Dsp} (\decay{\Lb}{\Dstarm \Lc}) decays with \decay{\Dsp}{\Km\Kp\pip} (\decay{\Lc}{\Km\proton\pip}) to a negligible level.
The mass of the \Km\pip\pip system from the \Dp candidate is recalculated using a kaon (proton) mass hypothesis for either of the pions.
The candidate is rejected if the pion has a high probability to be identified as a kaon (proton) and the recomputed mass is compatible with the known \Dsp (\Lc) mass \cite{PDG2020}.
Background contributions that arise from \decay{\Pphi(1020)}{\Km\Kp} transitions in the \Dsp decay chain are further suppressed by rejecting candidates if the pion has a high probability to be identified as a kaon and the mass of the \Km\pip system, where the kaon mass is assigned to either of the pions from the \Dp decay, is compatible with the known $\Pphi(1020)$ meson mass \cite{PDG2020}.
Decays of $\B^0_{(\squark)}$ mesons of the form \decay{\ensuremath{\B^0_{(\squark)}}}{\Dstarm h^- h^+ h^+} are suppressed by ensuring that the \Bsd and \Dp decay vertices are well separated.
Partially reconstructed decays, \ie decays where one or more final-state particles are not reconstructed, contribute to the lower-mass sideband and are accounted for in the fit to the data.

To suppress combinatorial background from random combinations of final-state tracks, a boosted decision tree~(BDT) classifier~\cite{Breiman,AdaBoost}, implemented in the TMVA toolkit~\cite{Hocker:2007ht,*TMVA4}, utilising the \mbox{AdaBoost} method is used.
The BDT classifier is trained using simulated \BtoDstD decays as signal proxy and the upper-mass sideband \mbox{($5450\mevcc<m_{\Dstarpm\Dmp}<6000\mevcc$)} as background proxy to avoid contributions from signal and partially reconstructed decays.
For each data-taking period a $k$-folding technique~\cite{Blum:1999:BHB:307400.307439} with $k=5$ is adopted.
The following variables are used in the training of the BDT classifier:
the mass difference of the \Dstarp and \Dz candidates; PID variables of the final-state particles of the \Dm candidate decay, the kaon coming from the \Dz candidate decay and the pion coming from the \Dstarp decay; the transverse momenta of the \Bsd candidate and the kaon from the \Dm decay; $t/\sigma_t$ of the \Dm candidate; the \chisqip of the \Bsd and \Dm candidates; the \chisq of the flight distance of the \Dm and \Dz candidates and the \chisq of a kinematic fit to the whole decay chain.

The optimal requirement on the BDT response (also referred to as working point) is determined by maximising the figure-of-merit $\varepsilon/(a/2+\sqrt{N_B})$~\cite{Punzi:2003bu}.
The efficiency of signal decays, $\varepsilon$, for a specific working point is determined by fits to the data around the known \Bd mass~\cite{PDG2020} before and after the application of the BDT requirement.
The number of background candidates in the \Bs signal region, $N_B$, is estimated with the upper-mass sideband, and the targeting significance in numbers of the standard deviation, $a$, is set to three.
A three-dimensional scan of the figure-of-merit in the three data-taking periods is conducted, resulting in slightly different working points.

Afterwards, a single candidate is selected randomly from each event containing multiple candidates.
The total selection efficiencies of the \BstoDstD and \BtoDstD decays are needed for the calculation of the branching fraction ratio and are calculated using simulated samples.

\section{Candidate mass fit}\label{sec:Massfit}
To improve the \Bds mass resolution, a kinematic fit is applied to the decay chain, where the masses of the \Dstarp, \Dz and \Dm candidates are constrained to their known values~\cite{PDG2020}.
An unbinned maximum-likelihood fit to the mass distribution of the \Dstarpm\Dmp system is performed separately for each data-taking period to determine the number of signal candidates.
To determine the significance of the observation of the \BstoDstD decay, the three likelihoods are added together.
The fit model consists of the signal \BtoDstD and \BstoDstD decays, a contribution from combinatorial background and components for partially reconstructed \decay{\Bd}{\Dstarpm\Dstarmp} and \decay{\Bs}{\Dstarpm\Dstarmp} decays, where one of the \Dstar mesons decays into a charged \D meson and an unreconstructed \piz meson or photon.
The \BtoDstD component is modelled by the sum of two Crystal Ball functions \cite{Skwarnicki:1986xj}, with the same mean but different widths and tail parameters.
The \BstoDstD component is described by the same model but with the mean shifted by the difference of the known \Bs and \Bd masses \cite{PDG2020}.
The parameters of the Crystal Ball functions are determined using fits to simulated \BtoDstD decays, apart from their mean and a single scale factor, which corrects the widths for inaccuracies in simulation.
The combinatorial background component is described by an exponential function.
The functional forms of the \decay{\Bd}{\Dstarpm\Dstarmp} and \decay{\Bs}{\Dstarpm\Dstarmp} contributions depend on the polarisation of the \Dstarpm mesons and are modelled using simulated decays with a combination of functions corresponding to pure longitudinal and transverse polarisations.
For a longitudinally polarised \Dstarpm meson the shape is a double peak, in contrast to a single broad peak for the case of a transversely polarised \Dstarpm system.
The free parameters in the fit are the mass of the \BtoDstD peak, the scaling factor, the slope of the exponential function, the relative fractions between longitudinally and transversely polarised \Dstarpm mesons in \decay{\Bd}{\Dstarpm\Dstarmp} and \decay{\Bs}{\Dstarpm\Dstarmp} decays, and the yields of all shapes.
Pseudoexperiments are used to validate that the model provides unbiased results.

\begin{figure}[tb]
\centering
\includegraphics[width=0.49\linewidth]{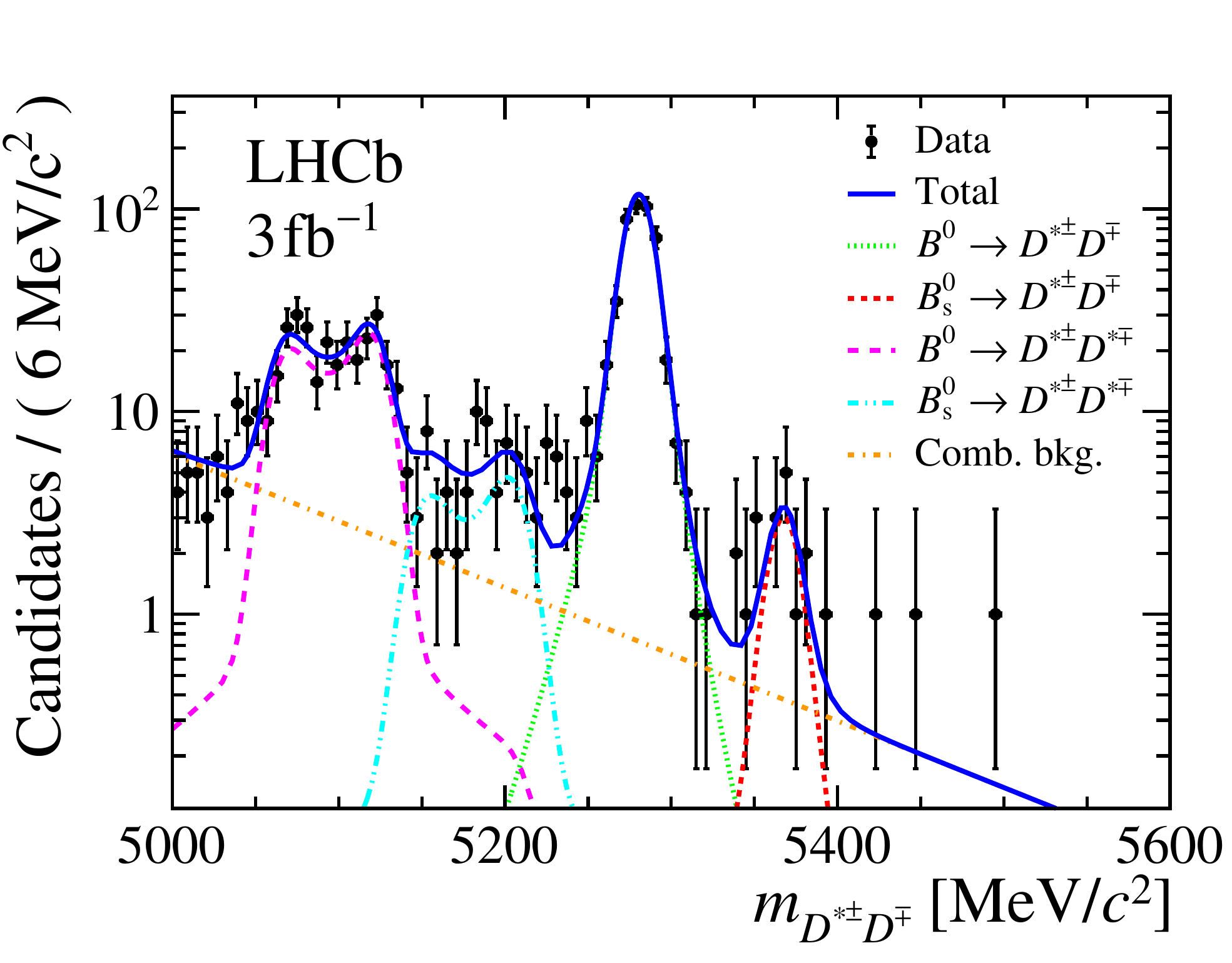}
\includegraphics[width=0.49\linewidth]{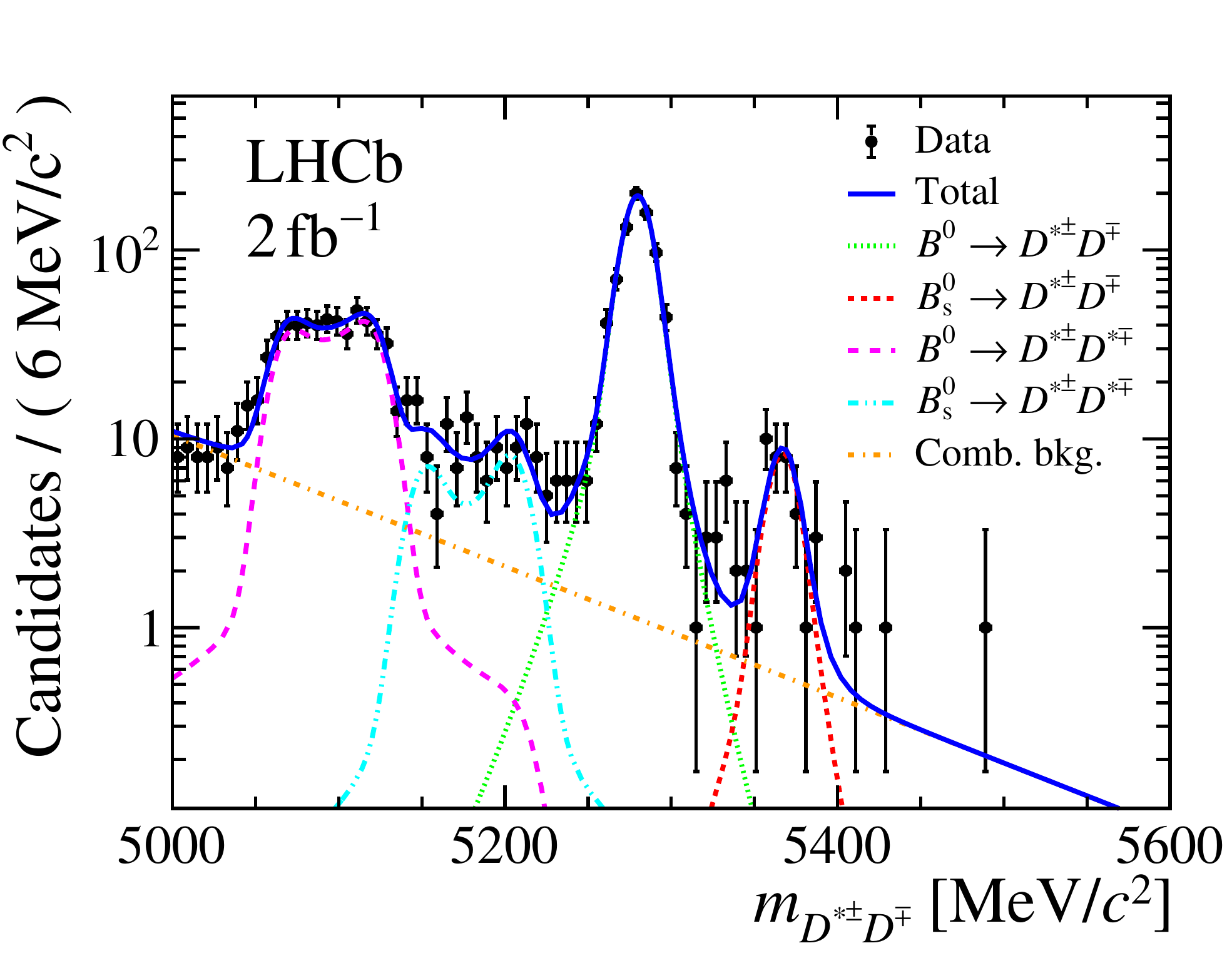}
\includegraphics[width=0.49\linewidth]{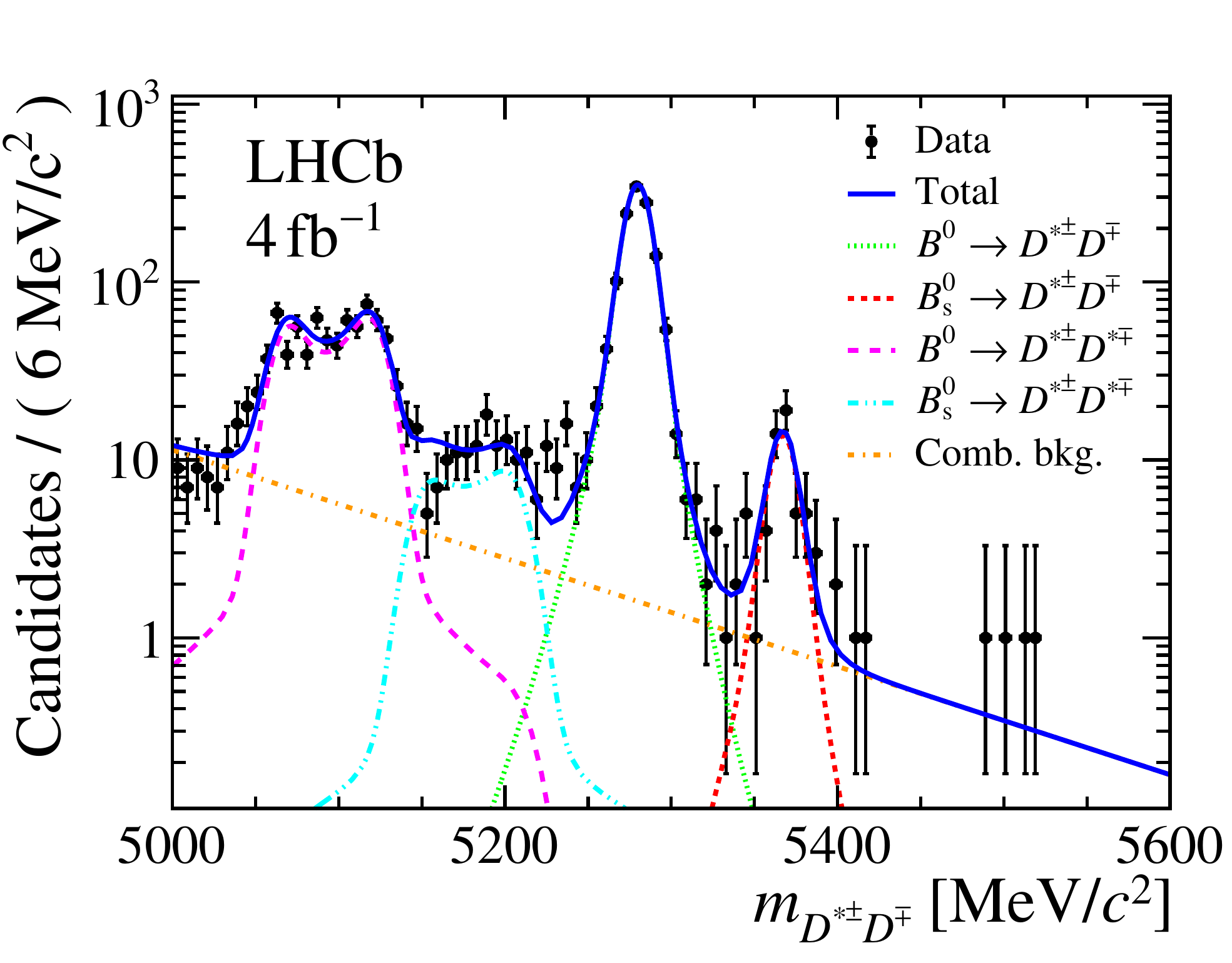}
\includegraphics[width=0.49\linewidth]{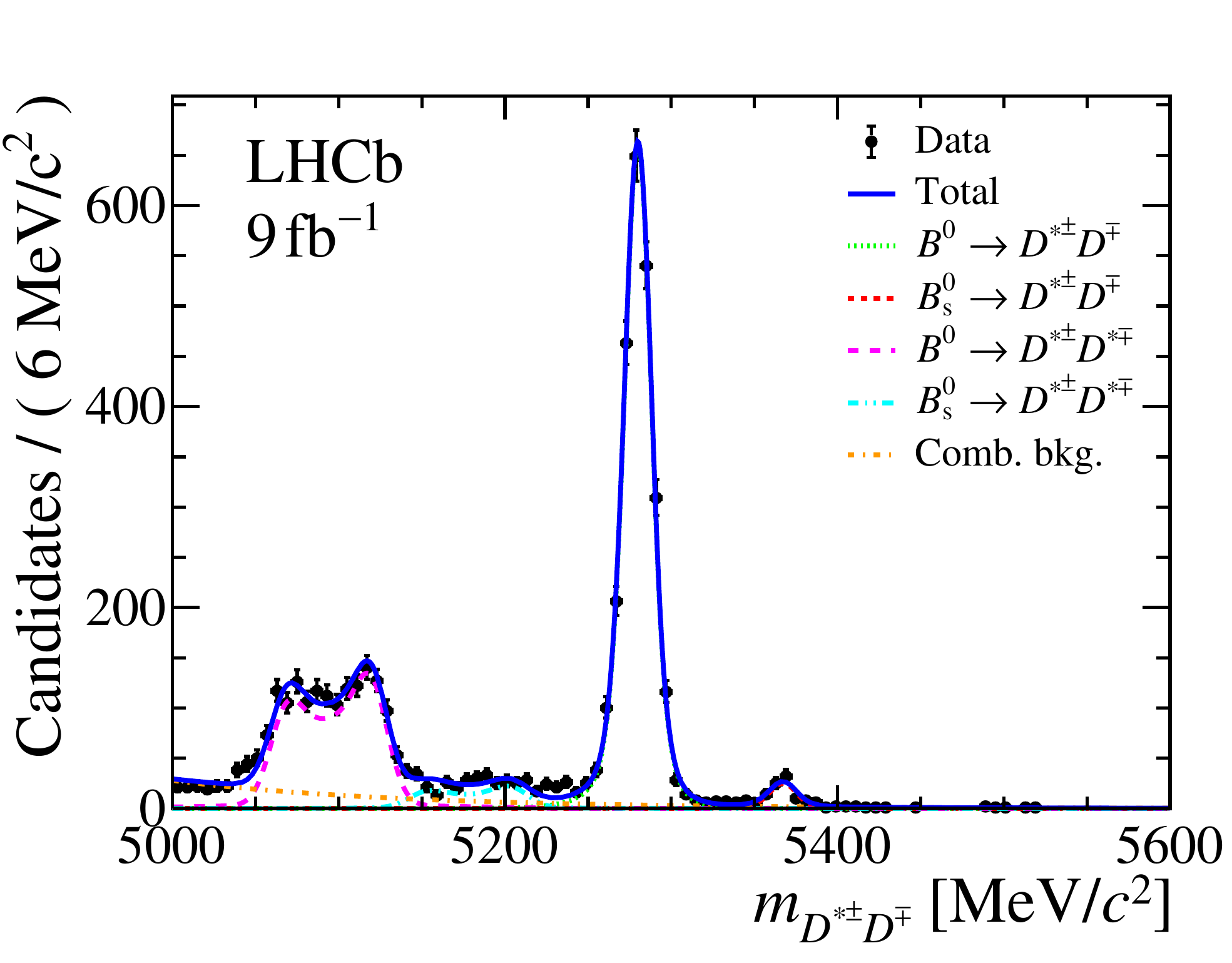}
\caption{The \Dstarpm\Dmp mass distributions for (top left) 2011--2012, (top right) 2015--2016, (bottom left) 2017--2018 data in logarithmic scale, and (bottom right) the combined data sample in linear scale. The total fit projection is shown as the blue solid line. The green dotted and the red dashed lines correspond to the signal contributions for the \Bd and \Bs decays, respectively. The orange dash-dotted line corresponds to the combinatorial background contribution. The \decay{\Bd}{\Dstarpm\Dstarmp} and \decay{\Bs}{\Dstarpm\Dstarmp} background components are described by the magenta long-dashed and the cyan long-dashed-two-dotted lines.}
\label{fig:selection:datafit}
\end{figure}

The resulting yields of \BtoDstD and \BstoDstD decays are $466\pm22$ and $12\pm 4$ in 2011--2012, $780\pm29$ and $34\pm 7$ in 2015--2016, and $1263\pm36$ and $49\pm 8$ in 2017--2018, respectively, where the quoted uncertainties are statistical only.
The resulting yields are checked by splitting the data in the two final states, \Dstarp\Dm and \Dstarm\Dp, and are found to be compatible.
The mass distributions and fit projections are shown in Fig.~\ref{fig:selection:datafit} for the three data-taking periods.

\section{Systematic uncertainties}\label{sec:Systematics}
The measurement of the ratio of branching fractions, $\BF(\BstoDstD)/\BF(\BtoDstD)$, relies on input from the measurement of the ratio of the \bquark quark hadronisation fractions to \Bs and \Bz mesons, $f_\squark/f_\dquark$.
The precision on $f_\squark/f_\dquark$ results in the dominant source of systematic uncertainty.
The values are taken from Ref.~\cite{fsfd} for 2011--2012 and from Ref.~\cite{LHCb-PAPER-2018-050} for 2015--2016 and 2017--2018 data-taking periods.
Both measurements share sources of systematic uncertainty and thus are treated as partially correlated.

Two sources of systematic uncertainty on the efficiency ratio are considered.
The first is caused by the finite size of the simulated data samples.
The second originates in the use of PID variables, whose distributions do not match perfectly between data and simulation.
This uncertainty is determined by choosing a different kernel density estimation in the transformation of the simulated PID response and calculating the difference of the resulting efficiency ratio.

The systematic uncertainties on the signal yields due to the fit models are evaluated using pseudoexperiments.
A systematic uncertainty due to the choice of the signal model and the assumption that the \Bd and \Bs distributions have identical shape in the mass fit is evaluated.
Candidates are generated with a mass distribution described by a Hypatia function~\cite{Santos:2013gra} with different parameters for the \Bd and \Bs models.
The values of the parameters of the Hypatia function are determined by a fit to simulated \BtoDstD and \BstoDstD data, respectively.
All yields and background parameters are set to the values found in the default fit to the data.
The generated candidates are then fitted with the default model and the result for the branching fraction ratio is calculated for each fit.
The mean of all experiments and its residual are calculated for the three periods separately.
The residual and its uncertainty are summed in quadrature and the square root is assigned as the systematic uncertainty.

In addition, a systematic uncertainty due to the model of the combinatorial background is evaluated.
Pseudoexperiments are used with parameters of the signal and partially reconstructed background models set to the values found in the default result.
The slope of the exponential function is extracted by a fit to the data, where a looser BDT requirement is applied to enhance the contribution of the combinatorial background.
The generated candidates are fitted with the default model and the systematic uncertainty is calculated in the same way as the systematic uncertainty for the signal model.

To determine the systematic uncertainty on the combined result, the systematic uncertainties related to the PID variables, signal model and background model are assumed to be fully correlated between the data-taking periods when calculating the weighted average.
The systematic uncertainties due to the finite size of the simulated data are assumed to be uncorrelated.
All systematic uncertainties are added in quadrature to obtain the total systematic uncertainty per data-taking period, and are listed together with their contributions in Table~\ref{tab:systematics}.
\begin{table}[tbp]
\centering
\caption{Systematic uncertainties on $\BF(\BstoDstD)/\BF(\BtoDstD)$. The systematic uncertainty is given relative to the measured value.}
\label{tab:systematics}
\begin{tabular}{lcccc}
\toprule
Source & 2011--2012 [\%] & 2015--2016 [\%] & 2017--2018 [\%] & Combined [\%] \\
\midrule
$f_s/f_d$ \cite{fsfd,LHCb-PAPER-2018-050} & 5.8 & 4.9 & 4.9 & 4.6\\
\midrule
Simulated data size & 0.8 & 1.2 & 0.8 & 0.6\\
PID & 0.7 & 0.7 & 0.8 & 0.7\\
Signal model & 0.1 & 0.1 & 0.5 & 0.3\\
Background model & 1.7 & 1.3 & 0.8 & 1.1\\
\midrule
Total without $f_s/f_d$ & 2.0 & 1.9 & 1.5 & 1.5\\
Total & 6.1 & 5.3 & 5.1 & 4.8\\
\bottomrule
\end{tabular}
\end{table}

\section{Results}\label{sec:Results}
The \BstoDstD decay is observed with a high significance, which is calculated using Wilks' theorem~\cite{Wilks:1938dza} together with the Neyman-Pearson lemma~\cite{Neyman:1933wgr}.
The relative branching ratio is calculated using the expression
\begin{align*}
\frac{\BF(\BstoDstD)}{\BF(\BtoDstD)} = \frac{N_\Bs}{N_\Bd} \frac{\varepsilon_\Bd}{\varepsilon_\Bs}\frac{f_\dquark}{f_\squark}\,,
\end{align*}
where the \Bs and \Bd yields, $N_\Bs$ and $N_\Bd$, are determined from the fit to the \Dstarpm\Dmp mass distribution.
The ratios of the \BstoDstD and \BtoDstD selection efficiencies, $\varepsilon_{\Bs}/\varepsilon_{\Bd}$, calculated using simulation samples for the data-taking periods 2011--2012, \mbox{2015--2016} and 2017--2018 are $1.063 \pm 0.008$, $1.062 \pm 0.013$ and $1.074 \pm 0.009$, respectively, where the uncertainties are statistical.
The ratios of the hadronisation fractions are taken as $0.259 \pm 0.015$ and $0.244 \pm 0.012$ for the 2011--2012~\cite{fsfd} and 2015--2018~\cite{LHCb-PAPER-2018-050} data-taking periods, respectively.

The ratios of branching fractions are found to be
\begin{align*}
\frac{\BF(\BstoDstD)}{\BF(\BtoDstD)}_{\text{2011--2012}} &= 0.093 \pm 0.032 \pm 0.002 \pm 0.005 \,, \\
\frac{\BF(\BstoDstD)}{\BF(\BtoDstD)}_{\text{2015--2016}} &= 0.168 \pm 0.034 \pm 0.003 \pm 0.008 \,, \\
\frac{\BF(\BstoDstD)}{\BF(\BtoDstD)}_{\text{2017--2018}} &= 0.149 \pm 0.024 \pm 0.002 \pm 0.007 \,,
\end{align*}
where the first uncertainty is statistical, the second systematic and the third is due to the uncertainty of the fragmentation fraction ratio $f_\squark/f_\dquark$.
Using the quadratic sums of the uncertainties as weights and including the correlation of the systematic uncertainties, the average of these measurements is
\begin{align*}
\frac{\mathcal{B}(B_s^0 \to D^{\ast \pm} D^{\mp}) }{\mathcal{B}(B^0 \to D^{\ast \pm} D^{\mp}) } = 0.137 \pm 0.017 \pm 0.002 \pm 0.006 \,.
\end{align*}

Using the measured value of the \BtoDstD branching fraction from Ref.~\cite{Rohrken:2012ta}, the \BstoDstD branching fraction is determined to be
\begin{align*}
\mathcal{B}(B_s^0 \to D^{\ast \pm} D^{\mp}) = \left(8.41 \pm 1.02 \pm 0.12 \pm 0.39 \pm 0.79 \right)\times 10^{-5} \,,
\end{align*}
where the fourth uncertainty is due to the \BtoDstD branching fraction.

This result assumes an average \Bs lifetime for the \BstoDstD decay.
The heavy and light eigenstates of the \Bs meson have significantly different lifetimes. As the selection efficiency depends on the lifetime, correction factors for the efficiency are calculated following the procedure outlined in Ref.~\cite{DeBruyn:2012wj} that considers either a purely heavy or a purely light \Bs eigenstate.
The correction factors are found to be compatible for all data-taking periods.
The integrated correction factors are 1.042 (0.949) for a purely heavy (light) \Bs eigenstate.
The equivalent effect in the selection efficiency of the \Bd decay is negligible due to the small value of \DGd~\cite{PDG2020}.

\section{Conclusion}\label{sec:Conclusion}

This paper presents the first observation of the \BstoDstD decay along with the measurement of its branching fraction relative to the \mbox{\BtoDstD} decay.
The analysis is performed with data collected by the LHCb experiment in the years 2011, 2012, and 2015 to 2018, corresponding to an integrated luminosity of $9\invfb$.
The combined ratio of branching fractions for all data-taking periods is determined to be
\begin{align*}
\frac{\mathcal{B}(B_s^0 \to D^{\ast \pm} D^{\mp}) }{\mathcal{B}(B^0 \to D^{\ast \pm} D^{\mp}) } = 0.137 \pm 0.017 \pm 0.002 \pm 0.006 \,,
\end{align*}
where the first uncertainty is statistical, the second systematic and the third is due to the uncertainty of the fragmentation fraction ratio $f_\squark/f_\dquark$.
The \BstoDstD branching fraction is determined to be
\begin{align*}
\mathcal{B}(B_s^0 \to D^{\ast \pm} D^{\mp}) = \left(8.41 \pm 1.02 \pm 0.12 \pm 0.39 \pm 0.79 \right)\times 10^{-5} \,,
\end{align*}
where the fourth uncertainty is due to the \BtoDstD branching fraction~\cite{PDG2020}.
The result is in agreement with predictions from other \B-meson decays \cite{Gronau:2012gs} and disagrees with predictions from a perturbative QCD approach~\cite{Li:2003az}.
It can be used to constrain subleading contributions in the measurement of the \CP-violating parameter $\sin(2\beta)$ with \BtoDstD decays.

%% file: acknowledgements.tex
\section*{Acknowledgements}
%
%
\noindent We express our gratitude to our colleagues in the CERN
accelerator departments for the excellent performance of the LHC. We
thank the technical and administrative staff at the LHCb
institutes.
We acknowledge support from CERN and from the national agencies:
CAPES, CNPq, FAPERJ and FINEP (Brazil); 
MOST and NSFC (China); 
CNRS/IN2P3 (France); 
BMBF, DFG and MPG (Germany); 
INFN (Italy); 
NWO (Netherlands); 
MNiSW and NCN (Poland); 
MEN/IFA (Romania); 
MSHE (Russia); 
MICINN (Spain); 
SNSF and SER (Switzerland); 
NASU (Ukraine); 
STFC (United Kingdom); 
DOE NP and NSF (USA).
We acknowledge the computing resources that are provided by CERN, IN2P3
(France), KIT and DESY (Germany), INFN (Italy), SURF (Netherlands),
PIC (Spain), GridPP (United Kingdom), RRCKI and Yandex
LLC (Russia), CSCS (Switzerland), IFIN-HH (Romania), CBPF (Brazil),
PL-GRID (Poland) and OSC (USA).
We are indebted to the communities behind the multiple open-source
software packages on which we depend.
Individual groups or members have received support from
AvH Foundation (Germany);
EPLANET, Marie Sk\l{}odowska-Curie Actions and ERC (European Union);
A*MIDEX, ANR, Labex P2IO and OCEVU, and R\'{e}gion Auvergne-Rh\^{o}ne-Alpes (France);
Key Research Program of Frontier Sciences of CAS, CAS PIFI, CAS CCEPP, 
Fundamental Research Funds for Central Universities, 
and Sci. \& Tech. Program of Guangzhou (China);
RFBR, RSF and Yandex LLC (Russia);
GVA, XuntaGal and GENCAT (Spain);
the Royal Society
and the Leverhulme Trust (United Kingdom).

%% file: LHCb_Authorship_22-Sep-2020.tex
\centerline
{\large\bf LHCb collaboration}
\begin
{flushleft}
\small
R.~Aaij$^{31}$,
C.~Abell{\'a}n~Beteta$^{49}$,
T.~Ackernley$^{59}$,
B.~Adeva$^{45}$,
M.~Adinolfi$^{53}$,
H.~Afsharnia$^{9}$,
C.A.~Aidala$^{84}$,
S.~Aiola$^{25}$,
Z.~Ajaltouni$^{9}$,
S.~Akar$^{64}$
% ...

{\footnotesize \it

% ...
$ ^{9}$Universit{\'e} Clermont Auvergne, CNRS/IN2P3, LPC, Clermont-Ferrand, France\\
% ...
$ ^{25}$INFN Sezione di Milano, Milano, Italy\\
% ...
$ ^{31}$Nikhef National Institute for Subatomic Physics, Amsterdam, Netherlands\\
% ...
$ ^{45}$Instituto Galego de F{\'\i}sica de Altas Enerx{\'\i}as (IGFAE), Universidade de Santiago de Compostela, Santiago de Compostela, Spain\\
% ...
$ ^{49}$Physik-Institut, Universit{\"a}t Z{\"u}rich, Z{\"u}rich, Switzerland\\
% ...
$ ^{53}$H.H. Wills Physics Laboratory, University of Bristol, Bristol, United Kingdom\\
% ...
$ ^{59}$Oliver Lodge Laboratory, University of Liverpool, Liverpool, United Kingdom\\
% ...
$ ^{64}$University of Cincinnati, Cincinnati, OH, United States\\
% ...
$ ^{67}$Syracuse University, Syracuse, NY, United States\\
% ...
$ ^{84}$University of Michigan, Ann Arbor, United States, associated to $^{67}$\\
\bigskip
% ...
\medskip
}
\end{flushleft}

%% file: main.bbl
\ifx\mcitethebibliography\mciteundefinedmacro
\PackageError{LHCb.bst}{mciteplus.sty has not been loaded}
{This bibstyle requires the use of the mciteplus package.}\fi
\providecommand{\href}[2]{#2}
\begin{mcitethebibliography}{10}
\mciteSetBstSublistMode{n}
\mciteSetBstMaxWidthForm{subitem}{\alph{mcitesubitemcount})}
\mciteSetBstSublistLabelBeginEnd{\mcitemaxwidthsubitemform\space}
{\relax}{\relax}

\bibitem{Cabibbo:1963yz}
N.~Cabibbo, \ifthenelse{\boolean{articletitles}}{\emph{{Unitary symmetry and
  leptonic decays}},
  }{}\href{https://doi.org/10.1103/PhysRevLett.10.531}{Phys.\ Rev.\ Lett.\
  \textbf{10} (1963) 531}\relax
\mciteBstWouldAddEndPuncttrue
\mciteSetBstMidEndSepPunct{\mcitedefaultmidpunct}
{\mcitedefaultendpunct}{\mcitedefaultseppunct}\relax
\EndOfBibitem
\bibitem{Kobayashi:1973fv}
M.~Kobayashi and T.~Maskawa,
  \ifthenelse{\boolean{articletitles}}{\emph{{\CP-violation in the
  renormalizable theory of weak interaction}},
  }{}\href{https://doi.org/10.1143/PTP.49.652}{Prog.\ Theor.\ Phys.\
  \textbf{49} (1973) 652}\relax
\mciteBstWouldAddEndPuncttrue
\mciteSetBstMidEndSepPunct{\mcitedefaultmidpunct}
{\mcitedefaultendpunct}{\mcitedefaultseppunct}\relax
\EndOfBibitem
\bibitem{Aubert:2006ia}
BaBar collaboration, B.~Aubert {\em et~al.},
  \ifthenelse{\boolean{articletitles}}{\emph{{Measurement of branching
  fractions and CP-violating charge asymmetries for B meson decays to $D^{(*)}
  \bar{D}^{(*)}$, and implications for the CKM angle $\gamma$}},
  }{}\href{https://doi.org/10.1103/PhysRevD.73.112004}{Phys.\ Rev.\
  \textbf{D73} (2006) 112004},
  \href{http://arxiv.org/abs/hep-ex/0604037}{{\normalfont\ttfamily
  arXiv:hep-ex/0604037}}\relax
\mciteBstWouldAddEndPuncttrue
\mciteSetBstMidEndSepPunct{\mcitedefaultmidpunct}
{\mcitedefaultendpunct}{\mcitedefaultseppunct}\relax
\EndOfBibitem
\bibitem{Aubert:2008ah}
BaBar collaboration, B.~Aubert {\em et~al.},
  \ifthenelse{\boolean{articletitles}}{\emph{{Measurements of time-dependent CP
  asymmetries in $B^0 \to D^{(*)+} D^{(*)-}$ decays}},
  }{}\href{https://doi.org/10.1103/PhysRevD.79.032002}{Phys.\ Rev.\
  \textbf{D79} (2009) 032002},
  \href{http://arxiv.org/abs/0808.1866}{{\normalfont\ttfamily
  arXiv:0808.1866}}\relax
\mciteBstWouldAddEndPuncttrue
\mciteSetBstMidEndSepPunct{\mcitedefaultmidpunct}
{\mcitedefaultendpunct}{\mcitedefaultseppunct}\relax
\EndOfBibitem
\bibitem{Fratina:2007zk}
Belle collaboration, S.~Fratina {\em et~al.},
  \ifthenelse{\boolean{articletitles}}{\emph{{Evidence for CP violation in $B^0
  \to D^+D^-$ decays}},
  }{}\href{https://doi.org/10.1103/PhysRevLett.98.221802}{Phys.\ Rev.\ Lett.\
  \textbf{98} (2007) 221802},
  \href{http://arxiv.org/abs/hep-ex/0702031}{{\normalfont\ttfamily
  arXiv:hep-ex/0702031}}\relax
\mciteBstWouldAddEndPuncttrue
\mciteSetBstMidEndSepPunct{\mcitedefaultmidpunct}
{\mcitedefaultendpunct}{\mcitedefaultseppunct}\relax
\EndOfBibitem
\bibitem{Rohrken:2012ta}
Belle collaboration, M.~Röhrken {\em et~al.},
  \ifthenelse{\boolean{articletitles}}{\emph{{Measurements of branching
  fractions and time-dependent CP violating asymmetries in $B^{0} \to
  D^{(*)\pm}D^{\mp}$ decays}},
  }{}\href{https://doi.org/10.1103/PhysRevD.85.091106}{Phys.\ Rev.\
  \textbf{D85} (2012) 091106},
  \href{http://arxiv.org/abs/1203.6647}{{\normalfont\ttfamily
  arXiv:1203.6647}}\relax
\mciteBstWouldAddEndPuncttrue
\mciteSetBstMidEndSepPunct{\mcitedefaultmidpunct}
{\mcitedefaultendpunct}{\mcitedefaultseppunct}\relax
\EndOfBibitem
\bibitem{LHCb-PAPER-2016-037}
LHCb collaboration, R.~Aaij {\em et~al.},
  \ifthenelse{\boolean{articletitles}}{\emph{{Measurement of \CP violation in
  \mbox{\decay{\B}{\Dp\Dm}} decays}},
  }{}\href{https://doi.org/10.1103/PhysRevLett.117.261801}{Phys.\ Rev.\ Lett.\
  \textbf{117} (2016) 261801},
  \href{http://arxiv.org/abs/1608.06620}{{\normalfont\ttfamily
  arXiv:1608.06620}}\relax
\mciteBstWouldAddEndPuncttrue
\mciteSetBstMidEndSepPunct{\mcitedefaultmidpunct}
{\mcitedefaultendpunct}{\mcitedefaultseppunct}\relax
\EndOfBibitem
\bibitem{LHCb-PAPER-2019-036}
LHCb collaboration, R.~Aaij {\em et~al.},
  \ifthenelse{\boolean{articletitles}}{\emph{{Measurement of \CP violation in
  \mbox{\decay{\Bz}{D^{*\pm} D^{\mp}}} decays}},
  }{}\href{https://doi.org/10.1007/JHEP03(2020)147}{JHEP \textbf{03} (2020)
  147}, \href{http://arxiv.org/abs/1912.03723}{{\normalfont\ttfamily
  arXiv:1912.03723}}\relax
\mciteBstWouldAddEndPuncttrue
\mciteSetBstMidEndSepPunct{\mcitedefaultmidpunct}
{\mcitedefaultendpunct}{\mcitedefaultseppunct}\relax
\EndOfBibitem
\bibitem{LHCb-PAPER-2014-051}
LHCb collaboration, R.~Aaij {\em et~al.},
  \ifthenelse{\boolean{articletitles}}{\emph{{Measurement of the \CP-violating
  phase \phis in \mbox{\decay{\Bsb}{\Dsp\Dsm}} decays}},
  }{}\href{https://doi.org/10.1103/PhysRevLett.113.211801}{Phys.\ Rev.\ Lett.\
  \textbf{113} (2014) 211801},
  \href{http://arxiv.org/abs/1409.4619}{{\normalfont\ttfamily
  arXiv:1409.4619}}\relax
\mciteBstWouldAddEndPuncttrue
\mciteSetBstMidEndSepPunct{\mcitedefaultmidpunct}
{\mcitedefaultendpunct}{\mcitedefaultseppunct}\relax
\EndOfBibitem
\bibitem{ALEKSAN1993173}
R.~Aleksan {\em et~al.}, \ifthenelse{\boolean{articletitles}}{\emph{{The decay
  $B \to D D^* + D^* D$ in the heavy quark limit and tests of CP violation}},
  }{}\href{https://doi.org/https://doi.org/10.1016/0370-2693(93)91588-E}{Phys.\
  Lett.\  \textbf{B317} (1993) 173 }\relax
\mciteBstWouldAddEndPuncttrue
\mciteSetBstMidEndSepPunct{\mcitedefaultmidpunct}
{\mcitedefaultendpunct}{\mcitedefaultseppunct}\relax
\EndOfBibitem
\bibitem{Sanda:1996pm}
A.~I. Sanda and Z.-z. Xing, \ifthenelse{\boolean{articletitles}}{\emph{{Towards
  determining $\phi_1$ with $B \to D^{(*)} \bar{D}^{(*)}$ }},
  }{}\href{https://doi.org/10.1103/PhysRevD.56.341}{Phys.\ Rev.\  \textbf{D56}
  (1997) 341}, \href{http://arxiv.org/abs/hep-ph/9702297}{{\normalfont\ttfamily
  arXiv:hep-ph/9702297}}\relax
\mciteBstWouldAddEndPuncttrue
\mciteSetBstMidEndSepPunct{\mcitedefaultmidpunct}
{\mcitedefaultendpunct}{\mcitedefaultseppunct}\relax
\EndOfBibitem
\bibitem{Xing:1998ca}
Z.-z. Xing, \ifthenelse{\boolean{articletitles}}{\emph{{Measuring CP violation
  and testing factorization in $B_d \to D^{*\pm} D^{\mp}$ and $B_s \to
  D_s^{*\pm} D_s^{\mp}$ decays }},
  }{}\href{https://doi.org/10.1016/S0370-2693(98)01285-4}{Phys.\ Lett.\
  \textbf{B443} (1998) 365},
  \href{http://arxiv.org/abs/hep-ph/9809496}{{\normalfont\ttfamily
  arXiv:hep-ph/9809496}}\relax
\mciteBstWouldAddEndPuncttrue
\mciteSetBstMidEndSepPunct{\mcitedefaultmidpunct}
{\mcitedefaultendpunct}{\mcitedefaultseppunct}\relax
\EndOfBibitem
\bibitem{Xing:1999yx}
Z.-z. Xing, \ifthenelse{\boolean{articletitles}}{\emph{{CP violation in $B_d
  \to D^+ D^-$, $D^{*+} D^-$, $D^+ D^{*-}$ and $D^{*+} D^{*-}$ decays}},
  }{}\href{https://doi.org/10.1103/PhysRevD.61.014010}{Phys.\ Rev.\
  \textbf{D61} (2000) 014010},
  \href{http://arxiv.org/abs/hep-ph/9907455}{{\normalfont\ttfamily
  arXiv:hep-ph/9907455}}\relax
\mciteBstWouldAddEndPuncttrue
\mciteSetBstMidEndSepPunct{\mcitedefaultmidpunct}
{\mcitedefaultendpunct}{\mcitedefaultseppunct}\relax
\EndOfBibitem
\bibitem{Pham:1999fy}
X.-Y. Pham and Z.-z. Xing, \ifthenelse{\boolean{articletitles}}{\emph{{CP
  asymmetries in $B_d \to D^{*+} D^{*-}$ and $B_s \to D_s^{*+} D_s^{*-}$
  decays: P wave dilution, penguin and rescattering effects}},
  }{}\href{https://doi.org/10.1016/S0370-2693(99)00611-5}{Phys.\ Lett.\
  \textbf{B458} (1999) 375},
  \href{http://arxiv.org/abs/hep-ph/9904360}{{\normalfont\ttfamily
  arXiv:hep-ph/9904360}}\relax
\mciteBstWouldAddEndPuncttrue
\mciteSetBstMidEndSepPunct{\mcitedefaultmidpunct}
{\mcitedefaultendpunct}{\mcitedefaultseppunct}\relax
\EndOfBibitem
\bibitem{Jung:2014jfa}
M.~Jung and S.~Schacht, \ifthenelse{\boolean{articletitles}}{\emph{{Standard
  model predictions and new physics sensitivity in $B \to DD$ decays}},
  }{}\href{https://doi.org/10.1103/PhysRevD.91.034027}{Phys.\ Rev.\
  \textbf{D91} (2015) 034027},
  \href{http://arxiv.org/abs/1410.8396}{{\normalfont\ttfamily
  arXiv:1410.8396}}\relax
\mciteBstWouldAddEndPuncttrue
\mciteSetBstMidEndSepPunct{\mcitedefaultmidpunct}
{\mcitedefaultendpunct}{\mcitedefaultseppunct}\relax
\EndOfBibitem
\bibitem{Bel:2015wha}
L.~Bel {\em et~al.}, \ifthenelse{\boolean{articletitles}}{\emph{{Anatomy of $
  B\to D\overline{D} $ decays}},
  }{}\href{https://doi.org/10.1007/JHEP07(2015)108}{JHEP \textbf{07} (2015)
  108}, \href{http://arxiv.org/abs/1505.01361}{{\normalfont\ttfamily
  arXiv:1505.01361}}\relax
\mciteBstWouldAddEndPuncttrue
\mciteSetBstMidEndSepPunct{\mcitedefaultmidpunct}
{\mcitedefaultendpunct}{\mcitedefaultseppunct}\relax
\EndOfBibitem
\bibitem{Gronau:2012gs}
M.~Gronau, D.~London, and J.~L. Rosner,
  \ifthenelse{\boolean{articletitles}}{\emph{Rescattering contributions to rare
  b-meson decays}, }{}\href{https://doi.org/10.1103/PhysRevD.87.036008}{Phys.\
  Rev.\  \textbf{D87} (2013) 036008},
  \href{http://arxiv.org/abs/1211.5785}{{\normalfont\ttfamily
  arXiv:1211.5785}}\relax
\mciteBstWouldAddEndPuncttrue
\mciteSetBstMidEndSepPunct{\mcitedefaultmidpunct}
{\mcitedefaultendpunct}{\mcitedefaultseppunct}\relax
\EndOfBibitem
\bibitem{Li:2003az}
Y.~Li, C.-D. Lu, and Z.-J. Xiao,
  \ifthenelse{\boolean{articletitles}}{\emph{{Rare decays
  \mbox{\decay{\Bz}{D_{s}^{(*)+} D_{s}^{(*)-}}} and \mbox{\decay{\Bz}{D^{(*)+}
  D^{(*)-}}} in perturbative QCD approach}},
  }{}\href{https://doi.org/10.1088/0954-3899/31/3/007}{J.\ Phys.\  \textbf{G31}
  (2005) 273}, \href{http://arxiv.org/abs/hep-ph/0308243}{{\normalfont\ttfamily
  arXiv:hep-ph/0308243}}\relax
\mciteBstWouldAddEndPuncttrue
\mciteSetBstMidEndSepPunct{\mcitedefaultmidpunct}
{\mcitedefaultendpunct}{\mcitedefaultseppunct}\relax
\EndOfBibitem
\bibitem{LHCb-DP-2008-001}
LHCb collaboration, A.~A. Alves~Jr.\ {\em et~al.},
  \ifthenelse{\boolean{articletitles}}{\emph{{The \lhcb detector at the LHC}},
  }{}\href{https://doi.org/10.1088/1748-0221/3/08/S08005}{JINST \textbf{3}
  (2008) S08005}\relax
\mciteBstWouldAddEndPuncttrue
\mciteSetBstMidEndSepPunct{\mcitedefaultmidpunct}
{\mcitedefaultendpunct}{\mcitedefaultseppunct}\relax
\EndOfBibitem
\bibitem{LHCb-DP-2014-002}
LHCb collaboration, R.~Aaij {\em et~al.},
  \ifthenelse{\boolean{articletitles}}{\emph{{LHCb detector performance}},
  }{}\href{https://doi.org/10.1142/S0217751X15300227}{Int.\ J.\ Mod.\ Phys.\
  \textbf{A30} (2015) 1530022},
  \href{http://arxiv.org/abs/1412.6352}{{\normalfont\ttfamily
  arXiv:1412.6352}}\relax
\mciteBstWouldAddEndPuncttrue
\mciteSetBstMidEndSepPunct{\mcitedefaultmidpunct}
{\mcitedefaultendpunct}{\mcitedefaultseppunct}\relax
\EndOfBibitem
\bibitem{Sjostrand:2007gs}
T.~Sj\"{o}strand, S.~Mrenna, and P.~Skands,
  \ifthenelse{\boolean{articletitles}}{\emph{{A brief introduction to PYTHIA
  8.1}}, }{}\href{https://doi.org/10.1016/j.cpc.2008.01.036}{Comput.\ Phys.\
  Commun.\  \textbf{178} (2008) 852},
  \href{http://arxiv.org/abs/0710.3820}{{\normalfont\ttfamily
  arXiv:0710.3820}}\relax
\mciteBstWouldAddEndPuncttrue
\mciteSetBstMidEndSepPunct{\mcitedefaultmidpunct}
{\mcitedefaultendpunct}{\mcitedefaultseppunct}\relax
\EndOfBibitem
\bibitem{Sjostrand:2006za}
T.~Sj\"{o}strand, S.~Mrenna, and P.~Skands,
  \ifthenelse{\boolean{articletitles}}{\emph{{PYTHIA 6.4 physics and manual}},
  }{}\href{https://doi.org/10.1088/1126-6708/2006/05/026}{JHEP \textbf{05}
  (2006) 026}, \href{http://arxiv.org/abs/hep-ph/0603175}{{\normalfont\ttfamily
  arXiv:hep-ph/0603175}}\relax
\mciteBstWouldAddEndPuncttrue
\mciteSetBstMidEndSepPunct{\mcitedefaultmidpunct}
{\mcitedefaultendpunct}{\mcitedefaultseppunct}\relax
\EndOfBibitem
\bibitem{LHCb-PROC-2010-056}
I.~Belyaev {\em et~al.}, \ifthenelse{\boolean{articletitles}}{\emph{{Handling
  of the generation of primary events in Gauss, the LHCb simulation
  framework}}, }{}\href{https://doi.org/10.1088/1742-6596/331/3/032047}{J.\
  Phys.\ Conf.\ Ser.\  \textbf{331} (2011) 032047}\relax
\mciteBstWouldAddEndPuncttrue
\mciteSetBstMidEndSepPunct{\mcitedefaultmidpunct}
{\mcitedefaultendpunct}{\mcitedefaultseppunct}\relax
\EndOfBibitem
\bibitem{Lange:2001uf}
D.~J. Lange, \ifthenelse{\boolean{articletitles}}{\emph{{The EvtGen particle
  decay simulation package}},
  }{}\href{https://doi.org/10.1016/S0168-9002(01)00089-4}{Nucl.\ Instrum.\
  Meth.\  \textbf{A462} (2001) 152}\relax
\mciteBstWouldAddEndPuncttrue
\mciteSetBstMidEndSepPunct{\mcitedefaultmidpunct}
{\mcitedefaultendpunct}{\mcitedefaultseppunct}\relax
\EndOfBibitem
\bibitem{Golonka:2005pn}
P.~Golonka and Z.~Was, \ifthenelse{\boolean{articletitles}}{\emph{{PHOTOS Monte
  Carlo: A precision tool for QED corrections in $Z$ and $W$ decays}},
  }{}\href{https://doi.org/10.1140/epjc/s2005-02396-4}{Eur.\ Phys.\ J.\
  \textbf{C45} (2006) 97},
  \href{http://arxiv.org/abs/hep-ph/0506026}{{\normalfont\ttfamily
  arXiv:hep-ph/0506026}}\relax
\mciteBstWouldAddEndPuncttrue
\mciteSetBstMidEndSepPunct{\mcitedefaultmidpunct}
{\mcitedefaultendpunct}{\mcitedefaultseppunct}\relax
\EndOfBibitem
\bibitem{Allison:2006ve}
Geant4 collaboration, J.~Allison {\em et~al.},
  \ifthenelse{\boolean{articletitles}}{\emph{{Geant4 developments and
  applications}}, }{}\href{https://doi.org/10.1109/TNS.2006.869826}{IEEE
  Trans.\ Nucl.\ Sci.\  \textbf{53} (2006) 270}\relax
\mciteBstWouldAddEndPuncttrue
\mciteSetBstMidEndSepPunct{\mcitedefaultmidpunct}
{\mcitedefaultendpunct}{\mcitedefaultseppunct}\relax
\EndOfBibitem
\bibitem{Agostinelli:2002hh}
Geant4 collaboration, S.~Agostinelli {\em et~al.},
  \ifthenelse{\boolean{articletitles}}{\emph{{Geant4: A simulation toolkit}},
  }{}\href{https://doi.org/10.1016/S0168-9002(03)01368-8}{Nucl.\ Instrum.\
  Meth.\  \textbf{A506} (2003) 250}\relax
\mciteBstWouldAddEndPuncttrue
\mciteSetBstMidEndSepPunct{\mcitedefaultmidpunct}
{\mcitedefaultendpunct}{\mcitedefaultseppunct}\relax
\EndOfBibitem
\bibitem{LHCb-PROC-2011-006}
M.~Clemencic {\em et~al.}, \ifthenelse{\boolean{articletitles}}{\emph{{The
  \lhcb simulation application, Gauss: Design, evolution and experience}},
  }{}\href{https://doi.org/10.1088/1742-6596/331/3/032023}{J.\ Phys.\ Conf.\
  Ser.\  \textbf{331} (2011) 032023}\relax
\mciteBstWouldAddEndPuncttrue
\mciteSetBstMidEndSepPunct{\mcitedefaultmidpunct}
{\mcitedefaultendpunct}{\mcitedefaultseppunct}\relax
\EndOfBibitem
\bibitem{Poluektov:2014rxa}
A.~Poluektov, \ifthenelse{\boolean{articletitles}}{\emph{{Kernel density
  estimation of a multidimensional efficiency profile}},
  }{}\href{https://doi.org/10.1088/1748-0221/10/02/P02011}{JINST \textbf{10}
  (2015) P02011}, \href{http://arxiv.org/abs/1411.5528}{{\normalfont\ttfamily
  arXiv:1411.5528}}\relax
\mciteBstWouldAddEndPuncttrue
\mciteSetBstMidEndSepPunct{\mcitedefaultmidpunct}
{\mcitedefaultendpunct}{\mcitedefaultseppunct}\relax
\EndOfBibitem
\bibitem{BBDT}
V.~V. Gligorov and M.~Williams,
  \ifthenelse{\boolean{articletitles}}{\emph{{Efficient, reliable and fast
  high-level triggering using a bonsai boosted decision tree}},
  }{}\href{https://doi.org/10.1088/1748-0221/8/02/P02013}{JINST \textbf{8}
  (2013) P02013}, \href{http://arxiv.org/abs/1210.6861}{{\normalfont\ttfamily
  arXiv:1210.6861}}\relax
\mciteBstWouldAddEndPuncttrue
\mciteSetBstMidEndSepPunct{\mcitedefaultmidpunct}
{\mcitedefaultendpunct}{\mcitedefaultseppunct}\relax
\EndOfBibitem
\bibitem{LHCb-PROC-2015-018}
T.~Likhomanenko {\em et~al.}, \ifthenelse{\boolean{articletitles}}{\emph{{LHCb
  topological trigger reoptimization}},
  }{}\href{https://doi.org/10.1088/1742-6596/664/8/082025}{J.\ Phys.\ Conf.\
  Ser.\  \textbf{664} (2015) 082025}\relax
\mciteBstWouldAddEndPuncttrue
\mciteSetBstMidEndSepPunct{\mcitedefaultmidpunct}
{\mcitedefaultendpunct}{\mcitedefaultseppunct}\relax
\EndOfBibitem
\bibitem{PDG2020}
Particle Data Group, P.~A. Zyla {\em et~al.},
  \ifthenelse{\boolean{articletitles}}{\emph{{\href{http://pdg.lbl.gov/}{Review
  of particle physics}}}, }{}\href{https://doi.org/10.1093/ptep/ptaa104}{Prog.\
  Theor.\ Exp.\ Phys.\  \textbf{2020} (2020) 083C01}\relax
\mciteBstWouldAddEndPuncttrue
\mciteSetBstMidEndSepPunct{\mcitedefaultmidpunct}
{\mcitedefaultendpunct}{\mcitedefaultseppunct}\relax
\EndOfBibitem
\bibitem{Breiman}
L.~Breiman, J.~H. Friedman, R.~A. Olshen, and C.~J. Stone, {\em Classification
  and regression trees}, Wadsworth international group, Belmont, California,
  USA, 1984\relax
\mciteBstWouldAddEndPuncttrue
\mciteSetBstMidEndSepPunct{\mcitedefaultmidpunct}
{\mcitedefaultendpunct}{\mcitedefaultseppunct}\relax
\EndOfBibitem
\bibitem{AdaBoost}
Y.~Freund and R.~E. Schapire, \ifthenelse{\boolean{articletitles}}{\emph{A
  decision-theoretic generalization of on-line learning and an application to
  boosting}, }{}\href{https://doi.org/10.1006/jcss.1997.1504}{J.\ Comput.\
  Syst.\ Sci.\  \textbf{55} (1997) 119}\relax
\mciteBstWouldAddEndPuncttrue
\mciteSetBstMidEndSepPunct{\mcitedefaultmidpunct}
{\mcitedefaultendpunct}{\mcitedefaultseppunct}\relax
\EndOfBibitem
\bibitem{Hocker:2007ht}
H.~Voss, A.~Hoecker, J.~Stelzer, and F.~Tegenfeldt,
  \ifthenelse{\boolean{articletitles}}{\emph{{TMVA - Toolkit for Multivariate
  Data Analysis with ROOT}}, }{}\href{https://doi.org/10.22323/1.050.0040}{PoS
  \textbf{ACAT} (2007) 040}\relax
\mciteBstWouldAddEndPuncttrue
\mciteSetBstMidEndSepPunct{\mcitedefaultmidpunct}
{\mcitedefaultendpunct}{\mcitedefaultseppunct}\relax
\EndOfBibitem
\bibitem{TMVA4}
A.~Hoecker {\em et~al.}, \ifthenelse{\boolean{articletitles}}{\emph{{TMVA 4 ---
  Toolkit for Multivariate Data Analysis with ROOT. Users Guide.}},
  }{}\href{http://arxiv.org/abs/physics/0703039}{{\normalfont\ttfamily
  arXiv:physics/0703039}}\relax
\mciteBstWouldAddEndPuncttrue
\mciteSetBstMidEndSepPunct{\mcitedefaultmidpunct}
{\mcitedefaultendpunct}{\mcitedefaultseppunct}\relax
\EndOfBibitem
\bibitem{Blum:1999:BHB:307400.307439}
A.~Blum, A.~Kalai, and J.~Langford,
  \ifthenelse{\boolean{articletitles}}{\emph{Beating the hold-out: bounds for
  k-fold and progressive cross-validation}, }{} in {\em Proceedings of the
  Twelfth Annual Conference on Computational Learning Theory},
  \href{https://doi.org/10.1145/307400.307439}{ COLT '99, (New York, NY, USA),
  203--208, ACM, 1999}\relax
\mciteBstWouldAddEndPuncttrue
\mciteSetBstMidEndSepPunct{\mcitedefaultmidpunct}
{\mcitedefaultendpunct}{\mcitedefaultseppunct}\relax
\EndOfBibitem
\bibitem{Punzi:2003bu}
G.~Punzi, \ifthenelse{\boolean{articletitles}}{\emph{{Sensitivity of searches
  for new signals and its optimization}}, }{}eConf \textbf{C030908} (2003)
  MODT002, \href{http://arxiv.org/abs/physics/0308063}{{\normalfont\ttfamily
  arXiv:physics/0308063}}\relax
\mciteBstWouldAddEndPuncttrue
\mciteSetBstMidEndSepPunct{\mcitedefaultmidpunct}
{\mcitedefaultendpunct}{\mcitedefaultseppunct}\relax
\EndOfBibitem
\bibitem{Skwarnicki:1986xj}
T.~Skwarnicki, {\em {A study of the radiative cascade transitions between the
  Upsilon-prime and Upsilon resonances}}, PhD thesis, Institute of Nuclear
  Physics, Krakow, 1986,
  {\href{http://inspirehep.net/record/230779/}{DESY-F31-86-02}}\relax
\mciteBstWouldAddEndPuncttrue
\mciteSetBstMidEndSepPunct{\mcitedefaultmidpunct}
{\mcitedefaultendpunct}{\mcitedefaultseppunct}\relax
\EndOfBibitem
\bibitem{fsfd}
LHCb collaboration, R.~Aaij {\em et~al.},
  \ifthenelse{\boolean{articletitles}}{\emph{{Measurement of the fragmentation
  fraction ratio $f_s/f_d$ and its dependence on $B$ meson kinematics}},
  }{}\href{https://doi.org/10.1007/JHEP04(2013)001}{JHEP \textbf{04} (2013)
  001}, \href{http://arxiv.org/abs/1301.5286}{{\normalfont\ttfamily
  arXiv:1301.5286}}, $f_s/f_d$ value updated in
  \href{https://cds.cern.ch/record/1559262}{LHCb-CONF-2013-011}\relax
\mciteBstWouldAddEndPuncttrue
\mciteSetBstMidEndSepPunct{\mcitedefaultmidpunct}
{\mcitedefaultendpunct}{\mcitedefaultseppunct}\relax
\EndOfBibitem
\bibitem{LHCb-PAPER-2018-050}
LHCb collaboration, R.~Aaij {\em et~al.},
  \ifthenelse{\boolean{articletitles}}{\emph{{Measurement of $b$-hadron
  fractions in 13\tev\ \proton\proton collisions}},
  }{}\href{https://doi.org/10.1103/PhysRevD.100.031102}{Phys.\ Rev.\
  \textbf{D100} (2019) 031102(R)},
  \href{http://arxiv.org/abs/1902.06794}{{\normalfont\ttfamily
  arXiv:1902.06794}}\relax
\mciteBstWouldAddEndPuncttrue
\mciteSetBstMidEndSepPunct{\mcitedefaultmidpunct}
{\mcitedefaultendpunct}{\mcitedefaultseppunct}\relax
\EndOfBibitem
\bibitem{Santos:2013gra}
D.~Mart{\'\i}nez~Santos and F.~Dupertuis,
  \ifthenelse{\boolean{articletitles}}{\emph{{Mass distributions marginalized
  over per-event errors}},
  }{}\href{https://doi.org/10.1016/j.nima.2014.06.081}{Nucl.\ Instrum.\ Meth.\
  \textbf{A764} (2014) 150},
  \href{http://arxiv.org/abs/1312.5000}{{\normalfont\ttfamily
  arXiv:1312.5000}}\relax
\mciteBstWouldAddEndPuncttrue
\mciteSetBstMidEndSepPunct{\mcitedefaultmidpunct}
{\mcitedefaultendpunct}{\mcitedefaultseppunct}\relax
\EndOfBibitem
\bibitem{Wilks:1938dza}
S.~S. Wilks, \ifthenelse{\boolean{articletitles}}{\emph{{The large-sample
  distribution of the likelihood ratio for testing composite hypotheses}},
  }{}\href{https://doi.org/10.1214/aoms/1177732360}{Ann.\ Math.\ Stat.\
  \textbf{9} (1938) 60}\relax
\mciteBstWouldAddEndPuncttrue
\mciteSetBstMidEndSepPunct{\mcitedefaultmidpunct}
{\mcitedefaultendpunct}{\mcitedefaultseppunct}\relax
\EndOfBibitem
\bibitem{Neyman:1933wgr}
J.~Neyman and E.~S. Pearson, \ifthenelse{\boolean{articletitles}}{\emph{{On the
  Problem of the Most Efficient Tests of Statistical Hypotheses}},
  }{}\href{https://doi.org/10.1098/rsta.1933.0009}{Phil.\ Trans.\ Roy.\ Soc.\
  Lond.\  \textbf{A231} (1933) 289}\relax
\mciteBstWouldAddEndPuncttrue
\mciteSetBstMidEndSepPunct{\mcitedefaultmidpunct}
{\mcitedefaultendpunct}{\mcitedefaultseppunct}\relax
\EndOfBibitem
\bibitem{DeBruyn:2012wj}
K.~De~Bruyn {\em et~al.}, \ifthenelse{\boolean{articletitles}}{\emph{{Branching
  ratio measurements of $B_s$ decays}},
  }{}\href{https://doi.org/10.1103/PhysRevD.86.014027}{Phys.\ Rev.\
  \textbf{D86} (2012) 014027},
  \href{http://arxiv.org/abs/1204.1735}{{\normalfont\ttfamily
  arXiv:1204.1735}}\relax
\mciteBstWouldAddEndPuncttrue
\mciteSetBstMidEndSepPunct{\mcitedefaultmidpunct}
{\mcitedefaultendpunct}{\mcitedefaultseppunct}\relax
\EndOfBibitem
\end{mcitethebibliography}
